\documentclass[a4paper]{jpconf}
\usepackage{graphicx}
\begin{document}
\title{Jet Physics in Heavy Ion Collisions with Compact Muon Solenoid detector
at the LHC}

\author{Igor Lokhtin (for the CMS Collaboration)}

\address{Skobeltsyn Institute of Nuclear Physics, Moscow State 
University, Moscow 119992, Russia}

\ead{Igor.Lokhtin@cern.ch}

\begin{abstract}
The status of CMS jet simulations and physics analysis in heavy ion collisions is 
presented. Jet reconstruction and high-p$_T$ particle tracking in the high 
multiplicity environment of heavy ion collisions at the LHC using the CMS 
calorimetry and tracking system are described. The Monte Carlo tools used to 
simulate jet quenching are discussed.
\end{abstract}

\section{Introduction}

One of the important tools for studying properties of the quark-gluon plasma (QGP) in
ultrarelativistic heavy ion collisions is QCD jet production. Medium-induced
energy loss of energetic partons, ``jet quenching'', should 
be very different in cold nuclear matter and in the QGP, resulting in many 
observable phenomena~\cite{baier_rev}. Recent RHIC data on 
high-p$_T$ particle production are in agreement with the jet quenching 
hypothesis~\cite{Wang:2004}. However direct event-by-event reconstruction of 
jets is not available in the RHIC experiments. At the LHC, a new regime of heavy ion 
physics will be reached at $\sqrt{s_{\rm NN}}=5.5$ TeV where hard and semi-hard 
particle production can dominate over the underlying soft events. The initial gluon 
densities in Pb + Pb reactions at the LHC are expected to be significantly higher 
than at RHIC, implying stronger partonic energy loss, observable 
in various new channels~\cite{Accardi:2003}. 

\section{CMS detector} 

The Compact Muon Solenoid (CMS) 
is a general purpose detector designed primarily to search for the Higgs boson 
in proton-proton collisions at the LHC~\cite{CMS:1994}. The detector is optimized 
for accurate measurements of the characteristics of high-energy leptons, 
photons and hadronic jets in a large acceptance, providing unique
capabilities for ``hard probes'' in both $pp$ and $AA$
collisions~\cite{Baur:2000}. A detailed description of the detector elements can 
be found in the corresponding Technical Design Reports~\cite{tdr1,tdr2,tdr3,tdr4}.
The central element of CMS is a $13$ m long, $6$ m 
diameter, high-field ($4$ T) solenoid with an internal radius of $\approx 3$ m. 
The tracker and muon chambers cover the pseudorapidity region $|\eta|<2.4$  
while the electromagnetic (ECAL) and hadron (HCAL) calorimeters reach 
$\eta= \pm 3$ and $\eta= \pm 5.2$ respectively. A pair of quartz-fibre 
very-forward (HF) calorimeters, located at $\pm 11$ m from the interaction 
point, cover the region $3<|\eta|<5.2$. In addition, the quartz-fibre 
calorimeter CASTOR covers the region $5.3<|\eta|<6.4$. The high precision 
tracker is composed of silicon pixel and strip counters and allows track 
momenta to be determined with a resolution better than $2$\% for tracks with 
$p_T$ between $0.5$ GeV/$c$ and a few tens of GeV/$c$.

\section{Jet production in heavy ion collisions at CMS}

The following probes for studying jet quenching with CMS has been 
proposed:

\noindent 
$\bullet$ high-$p_T$ jet pair production~\cite{Lokhtin:2000}; \\ 
$\bullet$ jets tagged by a leading charged hadron or neutral
pion~\cite{Lokhtin:2003}; \\    
$\bullet$ $B$-jets tagged by a leading muon~\cite{Lokhtin:2004epj2}; \\  
$\bullet$ jets produced opposite a gauge boson in $\gamma
+$jet~\cite{Wang:1996,Kodolova:1999} 
and $\gamma ^{*}/Z^0 +$jet~\cite{Kartvelishvili:1996,Awes:2003,Lokhtin:2004plb} 
final states; \\  
$\bullet$ high mass dimuons from semileptonic $B$ and $D$
decays~\cite{Lin:1999,Lokhtin:2001} and secondary $J/\psi$~\cite{Lokhtin:2001}; \\ 
$\bullet$ inclusive high-$p_T$ particle spectra~\cite{Gyulassy:1992}; \\ 
$\bullet$ energy flow measurements~\cite{Lokhtin:2002,Lokhtin:2004epj1}

\begin{table}[htb] 
\begin{center}
\begin{tabular}{|l|c|} \hline  
Channel & {\rm Time}=1.2$\times$10$^{6}$ s, $\sigma _{AA}=A^2\sigma _{pp}$ \\ \hline 
jet$+$jet, $E_T^{\rm jet}>100$ GeV & 4$\times$10$^6$ \\ \hline   
jet tagged by $h^{\pm}$/$\pi^0$, 
$E_T^{\rm jet}>100$ GeV, $z>0.5$ & 2$\times$10$^5$\\ \hline
$B$-jet tagged by $\mu$, $E_T^{\rm jet}>50$ GeV, $z>0.3$ & 2$\times$10$^4$\\ \hline
\end{tabular}
\caption{\label{tab1}Expected rates for some jet channels 
in a one month Pb + Pb run ($z=E_T^{\rm leader}/E_T^{\rm jet}$).}
\end{center}
\end{table}

The dependence of these probes on event centrality (determined 
using very forward CMS calorimetry~\cite{Damgov:2001}) and their azimuthal 
distributions (the event plane can be reconstructed using energy flow in 
CMS endcaps~\cite{Lokhtin:2003note}) will carry information about the  
properties of the QGP at the LHC. The event rates for some channels, including 
hard jets, in a one month Pb + Pb run (assuming two weeks of data taking and not 
including reconstruction and trigger efficiencies) in the CMS acceptance estimated 
with PYTHIA 6.2~\cite{pythia} are presented in \tref{tab1}.

\section{Jet reconstruction with calorimetry}

\begin{figure}[htbp]
\begin{minipage}{18pc}
\includegraphics[width=18pc]{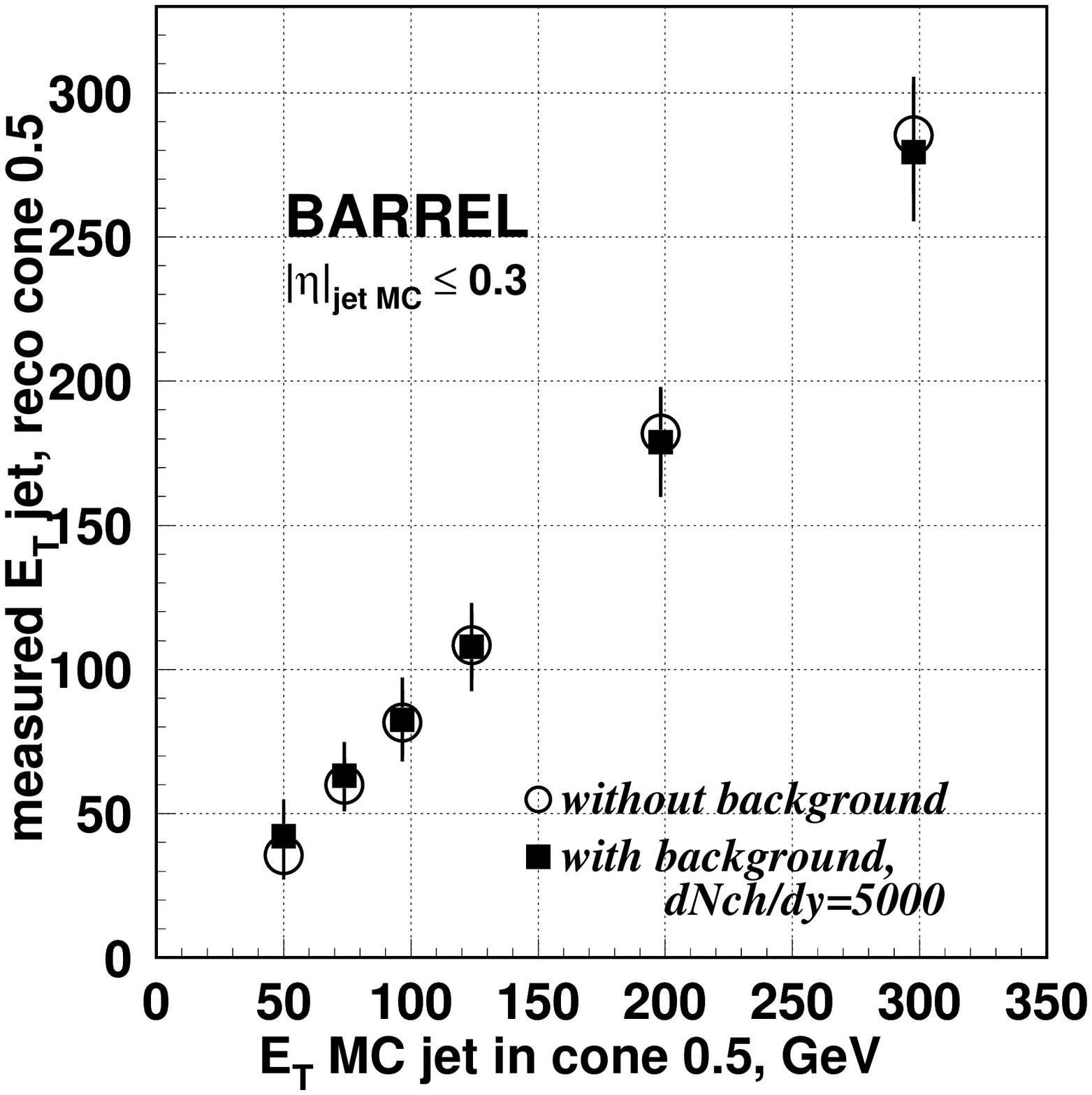}
\caption{\label{jet-en} Correlation between reconstructed and generated 
jet transverse energies.}
\end{minipage}\hspace{2pc}%
\begin{minipage}{18pc}
\includegraphics[width=18pc]{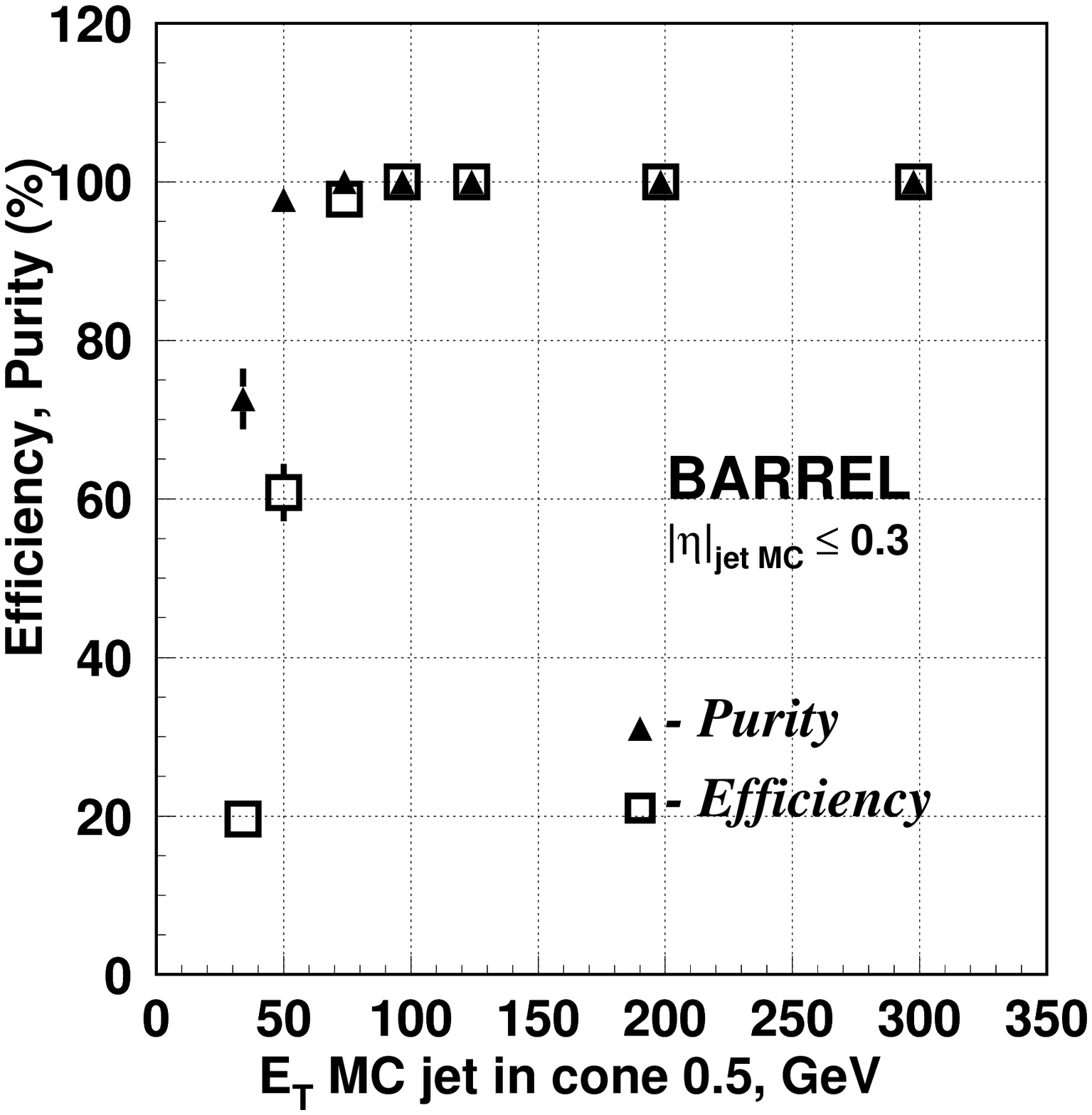}
\caption{\label{jet-ef} Purity and efficiency of jet reconstruction versus 
generated jet energy.}
\end{minipage} 
\end{figure}

The main difficulty of QCD jet recognition in heavy ion collisions arises from 
the ``false'' jet background -- transverse energy fluctuations coming from the 
high event multiplicity. In the CMS heavy ion programme, the sliding 
window-type jet finding algorithm has been
developed~\cite{Baur:2000,Accardi:2003} to search for ``jet-like'' 
clusters above the average energy and to subtract the $\eta$-dependent 
background from the underlying event. \Fref{jet-en} shows the linear correlation between 
reconstructed (full GEANT-based simulation) and generated (PYTHIA 6.2) 
transverse energies of jets with cone radius $R=0.5$ in Pb + Pb ($dN^{\pm}(y=0)/dy=5000$) 
and $pp$ events. The purity of jet reconstruction, defined as the number of 
events with a true QCD jet divided by the number of events with reconstructed 
jets, becomes $\sim 1$ at $100$ GeV (\fref{jet-ef}). Note also that the fine
angular resolution, $\Delta \varphi \sim \Delta \eta \sim 0.3$ at 
$E_T^{\rm MC jet}=100$ GeV, less than the azimuthal size of a 
calorimeter tower. Further development of the jet reconstruction
algorithm using tracker information for the jet energy correction is under way.  

\begin{figure}[htbp]
\begin{minipage}{18pc}
\includegraphics[width=18pc]{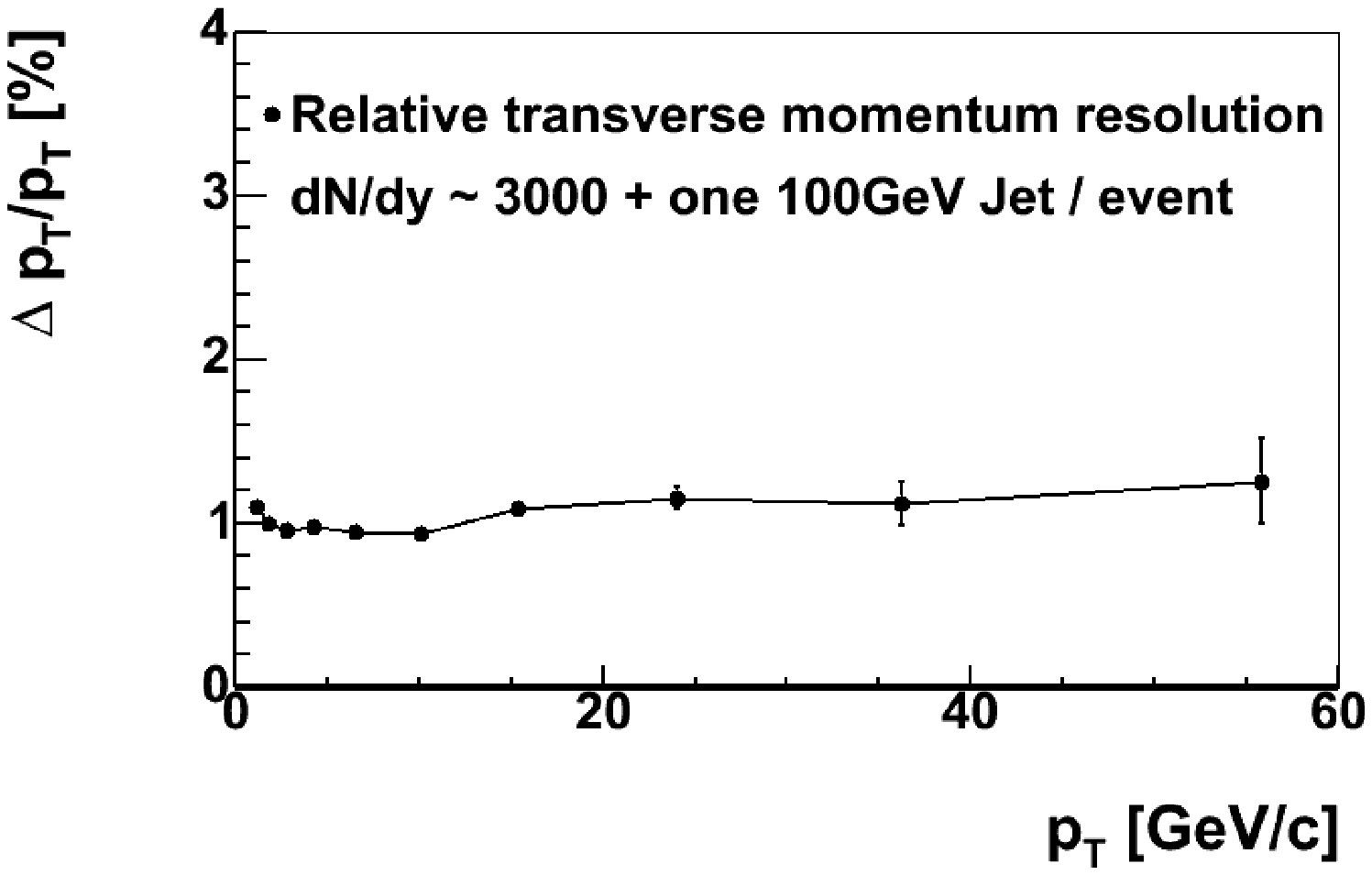}
\caption{\label{pt-res} Relative transverse momentum resolution of tracks 
inside a $100$ GeV jet.}
\end{minipage}\hspace{2pc}%
\begin{minipage}{18pc}
\includegraphics[width=18pc]{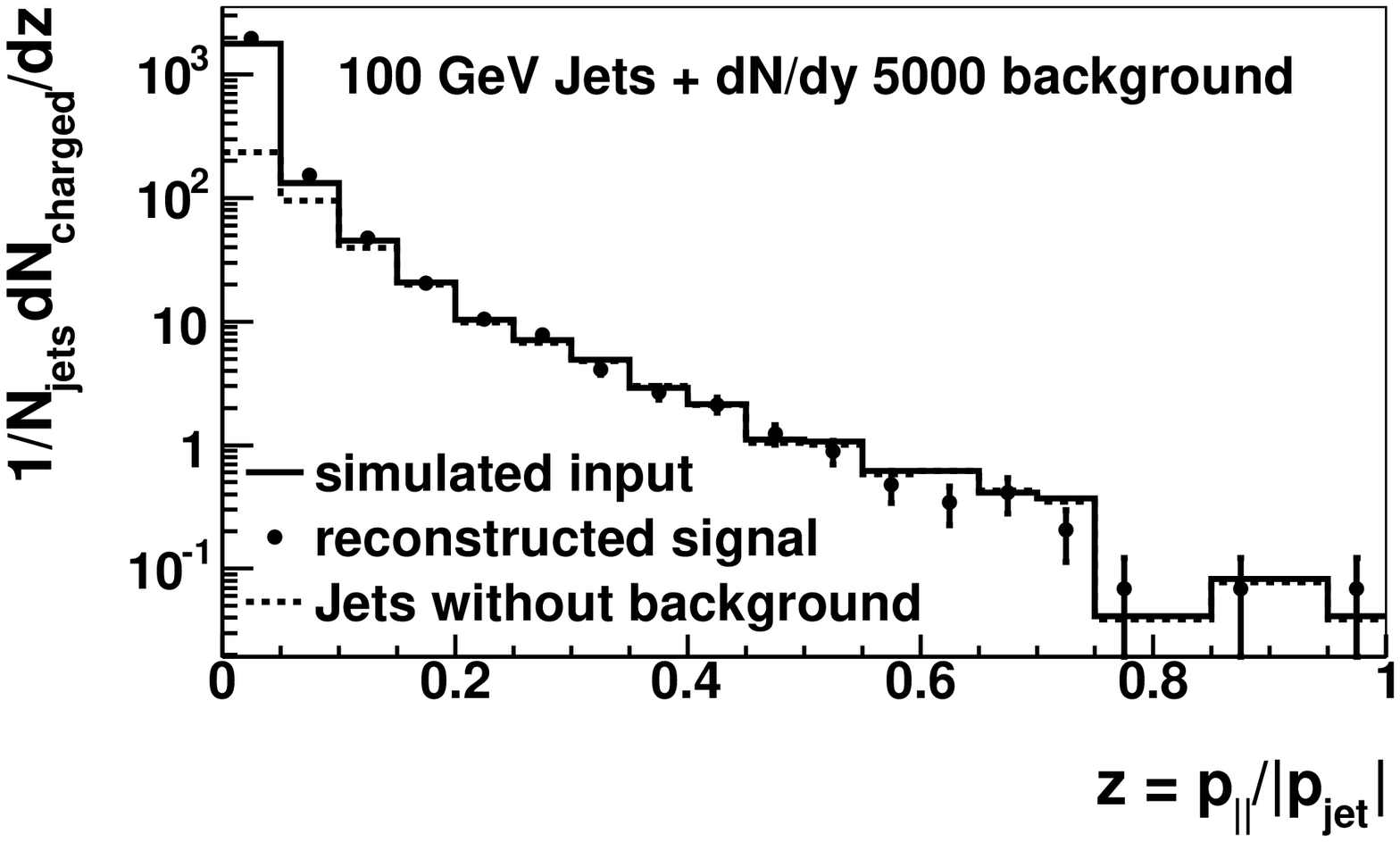}
\caption{\label{jetpt} Distribution over momentum fraction $z$ along the 
$100$ GeV jet axis.}
\end{minipage} 
\end{figure} 

\section{Charged particle tracking in jets} 

Track finding in heavy ion collisions is difficult due to the large number of 
tracks in an event. In addition to the primary tracks, the CMS tracker is 
occupied by secondaries produced by interactions with the detector material. 
The CMS track reconstruction algorithm, originally developed for $pp$ 
collisions, is based on Kalman Filtering and includes seed generation, track 
propagation, trajectory updating and smoothing. In order to reduce the 
combinatorial background during track seeding in heavy ion collisions, the 
modified track finder also includes primary vertex finding and restriction 
of the vertex region~\cite{Roland:2003}. The requirement for tracks to leave 
the tracker trough 
the outermost layer leads to a minimum transverse momentum cutoff of 
$p_T>1$ GeV/$c$ for the track to be considered reconstructable. Given this 
constraint, the track finder gives about $\approx 80$\% reconstruction efficiency 
and low fake rate even at high track densities of $dN^{\pm}(y=0)/dy=3000-5000$. 
The momentum resolution is less than $2$\% for $p_T<100$ GeV/$c$ 
(\fref{pt-res}). As an example, the reconstructed jet fragmentation
function for $E_T^{\rm jet}=100$ GeV using CMS tracker is shown in 
Figure \fref{jetpt}. 

\section{Monte-Carlo tools to simulate jet quenching} 

In most available event generators for LHC energies such important medium-induced 
effects such as jet quenching and elliptic flow are not included 
or are implemented poorly. Thus, in order to test the sensitivity of  
LHC observables to QGP formation, and to study the corresponding experimental 
capabilities of real detectors, the creation of adequate fast  
Monte-Carlo tools is necessary. One such tool developed recently to 
simulate rescattering and medium-induced partonic energy loss, is the fast 
event generator PYQUEN~\cite{pyquen}, implemented to modify a standard 
PYTHIA~\cite{pythia} jet event. The code HYDJET merges a fast generator of flow 
effects~\cite{hydro} with PYQUEN by simulating full heavy ion events as a superposition 
of soft, hydro-type state and hard multi-jets. These tools are very useful for detailed 
testing of the CMS capability to observe various probes of jet quenching and 
preparations for the heavy ion physics part of CMS Technical Design Report.
 
\section{Conclusions} 

With its large acceptance, nearly hermetic fine granularity hadronic and 
electromagnetic calorimetry, and good muon and tracking systems, CMS is an 
excellent device for the study of 
medium-induced energy loss by light and heavy quarks (``jet quenching'') at  
the LHC. Adequate jet reconstruction and high-$p_T$ particle tracking with CMS 
in the high multiplicity environment are possible. Significant progress in 
development of Monte-Carlo tools to simulate jet quenching in CMS has been 
achieved.

\section*{Acknowledgments}

The author wishes to express his gratitude to the members of CMS Collaboration, 
especially to Daniel Denegri, Christof Roland, Sergey Petrushanko, Ludmila 
Sarycheva, Alexander Snigirev, Constantin Teplov, Irina Vardanyan, Ramona Vogt and 
Boleslaw Wyslouch, for support and useful discussion. The author thanks the organizers 
of ICPAQGP 2005 for the warm welcome and hospitality. I also gratefully 
acknowledge partial support from Russian Foundation for Basic Research 
(grant N 04-02-16333).

\end{document}